\documentclass[12pt,a4paper]{article}
\usepackage[latin1]{inputenc}
\usepackage{amsmath}
\usepackage{amsfonts}
\usepackage{amssymb}

\newcommand{\hf}{\frac{1}{2}}

\newcommand{\e}{e^{i kX(z)}}

\newcommand{\p}{\partial}                                           
\newcommand{\pb}{\bar \partial}                                           
\newcommand{\pp}{\partial ^{2}}

\newcommand{\eps}{ \epsilon}                                        
\newcommand{\al}{\alpha }

\newcommand{\be}{\begin{equation}}
\newcommand{\br}{\begin{eqnarray}}
\newcommand{\ee}{\end{equation}} 
\newcommand{\er}{\end{eqnarray}}

\begin{document}
\title{
\hfill\parbox{3cm}{\normalsize IMSc/2009/02/02 \\ 
                               }\\
\vspace{1cm}
{\bf Gauge Invariant Action for the Open Bosonic String: Tachyon Action}
\author{B. Sathiapalan\\ {\em Institute of Mathematical Sciences}\\
{\em Taramani}\\{\em Chennai, India 600113}\\ bala@imsc.res.in}}
\maketitle       

\begin{abstract} 
A gauge invariant action for the open bosonic string has been proposed in an earlier paper. We work
out the consequences of this proposal for the lowest mode, viz. the tachyon.
 The action can  be calculated for generic momenta, perturbatively, order by order in the tachyon field. For on shell tachyons we explicitly calculate the cubic action and show that it reproduces the correct equations of motion and coincides wih the $\beta$ function to the required order. The calculation is done in terms of bare fields with a finite cutoff, which is the original prescription. We also show that it is possible in some momentum regions to renormalize the theory and eliminate the cutoff dependence so that the continuum limit can be taken. After renormalization, the parameter $R\over a$ is replaced by $R\over L$ where $R$ is an IR cutoff, $a$ is the UV cutoff and $L$ is some renormalization scale. There is also some arbitrariness in the overall normalization due to the choice of regularization scheme - this does not affect on-shell quantities. We also rederive within this scheme, the action in the region of zero momentum, which gives the exact (tree level) tachyon potential. The tachyon potential is consistent with Sen's conjecture that the height of the potential is the same as the tension of the brane.

\end{abstract}
\newpage

\section{Introduction}
 
The sigma model approach to obtaining the effective action for the fields of the string has primarily been used for massless fields \cite{L,CDMP,AS1,FT1,CG} although it was shown very early that it could be used for the massive modes (including closed string tachyon \cite{DS}). In \cite{BSPT} this technique was first applied to the open string and it was shown to all orders in perturbation theory that the equation of motion for the tachyon is proportional to the beta function, with the proportionality factor being the Zamolodchikov metric, something that was first argued on general grounds in \cite{Poly}. Subsequently equations for higher spin (spin one and higher) fields were considered and contact was attempted with string field theory \cite{HLP,BM,KPP,BSOS,ACNY,T,BSLV}. Interest in the tachyon has revived since the work on tachyon condensation in the open string \cite{AS2,AS3}. Different approaches have been used to understand this and also to prove the Sen conjectures. Cubic String Field Theory \cite{W} in particular has contributed significantly to this endeavour \cite{FK}.

    One parallel approach is the background independent formalism
proposed by Witten\cite{WiI,WiII,LW}. It seems well suited to the problem of tachyon potential (i.e zero momentum tachyons) as worked out in the original papers as well as in subsequent ones \cite{ShI,ShII,KMM}. The advantage of this method is that it uses the sigma model formalism and is thus manifestly background independent unlike string field theory. 
Also it is less tedious computationally. The space time geometry is manifest. The method seems to have some limitations \cite{LW,ShII} in dealing with on shell tachyons. These  have to do with the problem of world sheet divergences which seem to clash with BRST symmetry. The divergences do have a physical meaning - they correspond to contributions to the equations
of motion. 

The loop variable approach is also based on the sigma model and is therefore manifestly background independent. Gauge invariance is present as a spacetime symmetry principle and does not depend on world sheet properties. World sheet divergences therefore can be regulated and RG techniques can freely be used. A gauge invariant action has been proposed. If one takes the limit $R>>a$ (where $a$ is the UV cutoff and $R$ the IR cutoff) the proposal reduces, for the tachyon, and also for gauge fixed fields, to the same expression as that of Witten's background independent scheme. It is interesting that the same expression is obtained from a completely different (at least superficially) approach. However the way gauge invariance is implemented is different. Furthermore there seems to be no problem extending this to on-shell fields as outlined in \cite{BSA} and worked out in more detail here.
 
There is an arbitrariness in the RG procedure, which corresponds to off-shell field redefinitions. However this needs to be fixed in order to calculate for instance the height of the tachyon potential. Our approach for this problem is essentially the same as \cite{WiI,WiII,KMM,ShI} and establishes that the height of the potential is the same as the D-brane tension but does not evaluate the tension in terms of $g_s$ and $\al'$. This presumably needs some further calculation that fixes the remaining arbitrariness in the scheme. In Witten's approach the BRST  symmetry seems to fix this \cite{GS}.

This paper is organized as follows: In section 2 we give a brief review of the proposal for the action.
In section 3 we calculate the quadratic (i.e) kinetic term and also the lowest order $\beta$ function. This is done for generic momenta. For on shell fields more explicit results are obtained. These calculations are done first with bare fields (with a finite cutoff) and then after renormalization in terms of renormalized fields. The two give mutually consistent equations. In section 4 we give the cubic term in the action.
Here also the calculation is done in the bare theory as well as in the renormalized theory. The equation of motion obtained from the action is shown to coincide with the $\beta$ function. We also outline how the quartic and higher order terms work out as a fairly obvious extension of the cubic result. In section 5 for completeness we rederive the exact tree level potential of the constant tachyon (zero momentum) and fix also the leading derivative correction by considering the quadratic profile. Again it is done in the bare theory as well as the renormalized theory. This section reproduces earlier results based on Witten's work, but here, for uniformity with the rest of the paper, we work on the UHP with an IR cutoff, rather than on the unit disc.  Section 6 contains a summary and some conclusions.

\section{Gauge Invariant Action and Disk Partition Function}

Our proposal for the gauge invariant action \cite{BSA} starts with the following gauge fixed action:

\be     \label{action}
S = K{\p \over \p ln ~ a} \{ {Z \over \int dz \int dw \langle O(z) O(w) \rangle}\}
\ee

Here Z is the disk partition function with boundary interactions turned on.\footnote{It is assumed to be normalized to $1$ when the boundary interactions are set to zero. Thus  $Z\equiv {Z[\phi]\over Z[0]}$} These correspond to non trivial backgrounds for the open string fields. $O$ is a dimension-one (i.e. marginal) operator. The argument for this is as follows. 
For three point and higher functions, the fully integrated partition function with on-shell backgrounds turned on, has divergences due to the integral over the non-compact SL(2,R) (Mobius) group that is a symmetry of the on shell amplitudes.These are only present for three and higher point functions. Thus for three and higher point functions one expects a divergence of the form $\int dx \int _{-R}^{x-a}dy {1\over y^2} ln~{y\over a} $ multiplying the S-matrix - with on-shell poles removed due to  regularization.\footnote{This idea is an old one and has been used in many places. See for eg\cite{BSPT,FT1,T}.}  The denominator is  $\int dx \int _{-R}^{x-a}dy {1\over y^2} $. The numerator and denominater differ by terms proportional to $ln {R\over a}$.  Thus when we divide by the denominator and take the derivative w.r.t $ln~a$, this gives something proportional to the action that reproduces the tree level S-matrix. 

The denominator is just the integrated two point function of a dimension one operator (e.g.an on-shell tachyon). It is being used as an overall normalization factor.
The kinetic term comes from a two point function. If the particles are exactly on shell this is exactly of the form of the denominator. So the ratio is one and differentiation by $d\over d~ln~a$ gives zero. This is the rationale for dividing by a term of this form.  The kinetic term should vanish on-shell, and this construction guarantees that. When the particle is off shell the two point function gives a different result - with a $ln ~y$ correction proportional to $p^2-m^2$ and gives a contribution of the same form as the three and higher point functions. Thus including this contribution we obtain the full action.  The constant $K$ is an overall normalization that needs to be specified. 

One of the main motivations for constructing an action using the operation $d\over d~ln~a$ is that this can be easily made gauge invariant (under space-time gauge transformations) using the loop variable formalism \cite{BSLV,BSREV}. In this formalism $ln ~a$ is replaced first by $\sigma$ (Liouville mode) and subsequently by a dressed up version of $\sigma$ (called $\Sigma$ in \cite{BSLV,BSREV}) which is a function of $\sigma$ and many extra variables introduced to paramterize gauge transformation. In this formalism $d\over d\Sigma$
acting on the partition function (expressed in terms of loop variables) is automatically gauge invariant.\footnote{In this sense it is reminiscent of the BRST operator}. 
Thus this construction gives an action that reproduces the S-matrix, and is furthermore gauge invariant {\em off shell}, it is a good candidate for the gauge invariant action. We do not describe the loop variable construction here, since in the case of the tachyon (interacting with itself) there is no issue of gauge invariance.

The idea that the partition function is closely related to the effective action has been suspected for a long time\cite{FT1,T}. What we have given is one precise statement of the connection. As shown below, when $R>>a$ it reduces to a relation proposed by Witten in his work on background independent action for the open string\cite{WiI,WiII,LW,ShI}. It is interesting that this should be the case, since the rationale for that construction seems to be, at least superficially, quite different. When two different approaches give the same result it increases our confidence in the result.

Explicit calculation with on shell fields and also with zero momentum fields was outlined in \cite{BSA}. The aim of this paper is to work out the details. Many aspects of the exact renormalization group(RG) \cite{RG1,RG2,HS1,HS2} are probably relevant for a proper understanding of these issues. For instance one issue is that of a finite cutoff. It is true that starting with a bare theory on a lattice , one reaches a fixed point after only an infinte number of iterations. Therefore one can expect to have a finite cutoff for off shell fields. However since we are always an infinite number of rescalings away from the fixed point, one can further rescale the cutoff by an arbitrary large amount and thus approach the continuum arbitrarily closely. 
Another way of saying this is that on the exact unique RG trajectory the equations don't depend on the cutoff scale
\cite{RG1,RG2} (modulo terms that vanish (exponentially) as ${R\over a}\rightarrow \infty$ where $R$ is some IR cutoff length scale\footnote{This was demonstrated by explicit evaluation of the exact RG for this boundary field theory in \cite{BSERG}}.) so one can make it arbitrarily small. But actually this is true only for the exact RG. When we try to solve the equations  iteratively to obtain an equation for the marginal couplings, one can expect $ln{L\over a}$ terms. This has to be dealt with by a renormalization.
 This is done in this paper.

Finally it is worth briefly mentioning the connection between the notion of world sheet Renormalization Group as well as the notion of renormalization, with the space time action and space time fields. The exact RG equations can be thought of as the exact (cubic) string field equations. The exact RG is quadratic as is the string field equation. When we solve for say, the massive fields in terms of massless fields we get a higher order equation for massless fields. This is like writing down the 
beta functions for marginal couplings by eliminating the irrelevant ones from the exact RG.\footnote{See \cite{Wil} for a clear description of this.} This is done iteratively and each iteration increases the degree of nonlinearity of the equation. This can be roughly compared with the level truncation in string field theory. As we iterate the RG and move towards the IR, it is equivalent to increases the level to arbitrary high values. When there are numerical coefficients in the equation that are of order $N$, the number of iterations, this shows up as factors of $ln {L\over a}$ in the relation between bare and renormalized couplings, where $a$ is the cutoff and $L$ is the renormalization scale. If we want to take $a$ to zero we need to do an infinite renormalization. 

\section{Action for Tachyon Felds with Generic Momenta: Quadratic Terms}

We shall discuss this first in terms of bare fields and then in terms of renormalized fields.

\subsection{Bare Fields}

\subsubsection{Partition Function,Z}
\be
Z = \langle e^{-\int _{\p \Gamma} {dz \over a} \phi _B (X(z))}\rangle
\ee  
We can write $\phi _B(X(z)) = \int dk~\phi _B(k)\e$. The subscript "B" denotes that it is a bare coupling 
constant of the two dimensional theory. $a$ is the world sheet cutoff.

The expectation value is calculated using the Polyakov measure
\[
\int {\cal D} X e ^{- {1\over \al '}\int _{\Gamma} d^2z \p X \pb X }
\]

We are working with a Euclidean metric on the world sheet. $\Gamma$ is
the upper half plane (UHP) because we are interested in the open
string and $\p \Gamma$ is the real axis. With this
normalization $\langle X(z) X(w) \rangle = - {\al '\over 2\pi} (ln ~ |{z-w\over 2R}|
+ ln ~ |{z-\bar w\over 2R}|)$ for the upper half plane with Neumann boundary
conditions on the real axis. The range of integration along the real axis is taken to be from $-R$ to $+R$. Thus $2R$ acts as an IR cutoff in the theory. Further, for a tachyon vertex operator
on the real axis at $z=x$, $e^{ik.X(x)} = :e^{ik.X(x)}: e^{{\al '
    k^2 \over 2\pi }ln ~ (a/2R)}$. Thus if we set $\al '=\pi$, $k^2=2$ 
ensures that the operator is marginal and this is
the mass shell condition for the open string tachyon. 

Let us evaluate $Z$ in powers of $\phi _B$.
\be   
Z~=~\langle 1 \rangle ~+~\langle-\int _{-R}^{+R} {dz \over a} \phi _B
(X(z))    \rangle  ~+~\langle \int _{-R+a}^{+R}  {dz_1 \over a}  
\phi _B (X(z_1))\int _{-R}^{z_1-a} {dz_2 \over a} \phi _B (X(z_2))  \rangle +...
\ee

Note that the lower end in one of the integrations is $-R+a$. This ensures that the two vertex operators are always separated by a minimum distnce of $a$. This is sharp cutoff and is not good for an exact RG treatment because it does not provide a clean separation between modes in momentum space. However if we are near on-shell and $R/a >>1$ this is not important. It has the advantage of being easy to calculate with. A better cutoff would be to use a cutoff propagator, eg. ${1\over 2\pi}ln~[(z-w)^2+a^2]$. 
Yet another procedure is to use a momentum space cutoff: $G(k) = {e^{-\eps k^2}\over k^2}$ as used in 
\cite{BSERG}. One can also cut off the momentum space integrals and we will use this last method in deriving the tachyon potential.
 
Assume for the moment that $\phi _B$ has non-zero space-time momentum so
that it is of the form $\int dk~\phi _B (k) \e$ with $\phi _B (0) =0$.
Then the linear term vanishes by momentum conservation. The quadratic
term is 
\[ 
\int dp ~\int dk~ \int _{-R+a}^{R} {dz \over a} ~\int _{-R} ^{z-a}
{du \over a} ({z-u\over 2R})^{k.p}\phi _B (k)  \phi _B (p) \delta (k+p)({a\over 2R})^{k^2 +p^2\over 2}
 \]
\be   \label{3.4}
=~\int dk \int _{-R+a}^{R} {dz \over a} ~\int   _{-R} ^{z-a}{du \over a}
({z-u\over 2R})^{-k^2}({a\over 2R}) ^{k^2} \phi _B (k)  \phi _B (-k)
\ee
 Doing the $u$ integral:
\be   \label{3.5}
=~-\int dk \int _{-R+a}^{R} dz 
{(z-u)^{-k^2+1}\over -k^2+1}\Big \|^{z-a} _{-R}(a) ^{k^2-2} \phi _B (k)  \phi _B (-k)
\ee
And the $z$-integral gives 
\be   \label{3.6}
-\int dk~{\{ {2R-a\over a} - {[({2R\over a})^{-k^2+2} -1]\over -k^2+2} \}\over -k^2+1} \phi _B (k)  \phi _B (-k)
\ee

\subsubsection{\bf \boldmath $\beta$ Beta-function:}

 We can now pause and calculate the $\beta$- function to second order. One method is to use
\be  \label{3.7}
{dZ\over dln~a}=0 = {\p Z\over \p ln~a} - \int dk~\beta _{\phi _B(k)} {\p Z \over \p \phi _B(k)}
\ee

To lowest order we have 
\be
\langle ~~\int {dx \over a} ~\phi _B (k) e^{{k^2\over 2} ln~{a\over 2R}}:\e :~~\rangle
\ee 

This gives:
\be
\beta _{\phi _B(k)}=-{d\phi _B (k)\over d ln~a} = ({k^2\over 2}-1)\phi _B (k)
\ee
We have ignored for the moment the fact that momentum conservation forces $k=0$. (This lowest order result for $\beta$ can be obtained by other methods as well.)
So the above method
only involves $\beta _{\phi _B (0)}$ to this order.

Keeping terms to the next order (from (\ref{3.6})):

\be
Z= -{2R\over a} \phi _B (0) - [{ ({2R\over a}-1) - {(({2R\over a})^{-k^2+2} -1)\over -k^2+2}\over -k^2+1}]\phi _B(k)\phi _B (-k)
\ee

We write $\beta _{\phi _B} = \beta ^{(1)}_{\phi _B} + \beta ^{(2)}_{\phi _B}= ({k^2\over 2}-1)\phi _{B} (k)+\beta ^{(2)}_{\phi _B} $ as the sum of the first order (linear in $\phi$) and the second order (quadratic in $\phi$) pieces. Then we get:
\[
{dZ\over dln~a}=0 \Rightarrow {2R\over a}\beta _{\phi _B(0)}+ {2R\over a}\phi _B (0) +
\]
\[
{-1\over -k^2+1}\{ 
[-{2R\over a} + ({2R\over a})^{-k^2+2}]\phi _B (k) \phi _B (-k)
\]
\be
 [({2R\over a}-1) -{({2R\over a})^{-k^2+2}-1\over -k^2+2}][2-k^2]\} (\phi _B (k) \phi _B (-k)) =0
\ee
\be   \label{betabare}
\Rightarrow  \beta _{\phi _B(0)} =-\phi _B (0) +\int dk~\{(1-     {a\over 2R}) (\phi _B (k) \phi _B (-k))  \}
\ee

{\boldmath $k^2=2$:}

When $k^2 = 2$ one has to be careful because of the pole. 
\be
Z = -{2R\over a} \phi _B (0) + [{2R\over a} -1-ln~{2R\over a}]\phi _B (k) \phi _B (-k)
\ee 

Setting ${dZ\over d ln~a}=0$ as before, we get the same expression for the beta function.  

\subsubsection{\bf Action:}

Now we turn to the action. Consider (\ref{3.6}):

This expression is valid for general $k$. If we now specialize to $k^2-2 \approx 0$ we get
\be
\int dk~\{ {2R\over a} -1 - ln~({2R\over a}) - {1\over 2} (-k^2+2) ln^2 ({2R\over a})\}\phi _B (k)  \phi _B (-k)
\ee
Setting $k^2-2=0$ in the above expression gives us the denominator of the formula (\ref{action}) for the Action.
According to (\ref{action}) the action,S, is thus given by
\be  \label{S}
 S~=~K{\p \over \p~ln~a}{-\int dk~\{ {2R\over a} -1 - ln~({2R\over a}) - {1\over 2} (-k^2+2) ln^2 ({2R\over a})\}\phi _B (k)  \phi _B (-k)\over
  {2R\over a} -1 - ln~({2R\over a}) }           
\ee
Thus we can write, for near on shell tachyons: 
\be
S (k^2\approx 2)~=~ K h(R/a) \int dk~(2-k^2) \phi _B (k) \phi _B (-k)
\ee
where 
\be  \label{h}
h(R/a) =  {\p \over \p~ln~a} [ {ln^2~({2R\over a})\over {2R\over a}-1 - ln~({2R\over a})}]
\ee

Thus even for near on-shell fields one cannot take $a \rightarrow 0$, the continuum limit, because $h$ vanishes.
This may seem surprising. But there is no real need to take the continuum limit. We know that for off shell fields we need a finite cutoff. So we can treat $R\over a$ as a parameter of the theory. What is required is that the
S-matrix should not depend on this parameter. We will see later, that the coefficient of the cubic term is also
exactly the same function. Thus we can absorb this ill defined factor into the overall factor $K$. The parameter $R\over a$ will still show up in off-shell vertices. This is thus a parameter of the off-shell theory that does not affect the on-shell theory.  This parameter is analogous to the number $4\over 3\sqrt{3}$ in cubic string field theory.\footnote{This is explained for instance in \cite{BSFC}}

The more general expression for $S$ is:

\be   \label{S_B}
S~=~K
{\p \over \p~ln~a} [{
-\int dk~{\{ {2R\over a} -1- {[({2R\over a})^{-k^2+2} -1]\over -k^2+2} \}\over -k^2+1} \phi _B (k)  \phi _B (-k)
\over {2R\over a} -1 - ln~({2R\over a}) }] 
\ee

In the limit $R>>a$ we can approximate the denominator, which is $\int dz~\int dw~\langle O(z)O(w)\rangle$
of (\ref{action}), by $2R\over a$. Then the expression for the action is
\be   \label{Sbeta}
S = K {\p \over \p ln~a}[{a\over 2r} Z] = K {a\over 2r}(1+\int dk~\beta _{\phi_B (k)}{\p \over \p \phi _B (k)})Z
\ee  
where we have used (\ref{3.7}).

This is a useful form and can be applied to the partition function written in terms of renormalized variables.
$\beta _{\phi _B}$ will be replaced by $\beta _{\phi _R}$. In this form it is the same as the expression derived by Witten (\cite{WiI,WiII}) where the vector field is identified with the beta function (\cite{ShI}).

{\bf Zero Momentum Tachyon:}

This expression is valid in particular for $k^2=0$, when it reduces to
\be   \label{S(0)}
K{\p \over \p ~ln ~ a}{
[{-({2R\over a}) +{1\over 2} +{1\over 2} ({2R\over a})^2\over {2R\over a} -1 - ln~({2R\over a})} ] \phi _B (k) \phi _B (-k)}
\ee

At zero momentum the tachyon coupling in the boundary action reduces to \[\Delta S_{world~sheet ~boundary} = \int _R^R~{dz\over a} \phi _B(0) = {2R\over a} \phi _B (0)\]. The contribution to the partition function is expected to be 
\be  \label{naive}
e^{-{2R\over a} \phi _B(0)}
\ee
 Consequently the quadratic term is expected to be $ {1\over 2} ({2R\over a})^2 \phi _B(0)^2$. This does not quite match with the numerator of (\ref{S(0)}). The discrepancy is due to the minimum distance of $a$ between vertex operators not being imposed in (\ref{naive}). If we do impose that rule one gets
\be 
\int _{-R+a}^R {dz \over a} \int _R ^{z-a} {dw\over a}  \phi _B (0)^2 = \hf (({2R\over a}) -1)^2 \phi _B (0)^2
\ee
 which agrees with the numerator.

In the limit $R/a \rightarrow \infty$ this difference does not matter. Furthermore this would not be an issue in some other schemes such as when we use a cutoff propagator.

{\bf \boldmath Tachyon at $k^2\approx 0$}

One can turn now to the order $k^2$ term in an expansion about zero momentum. We consider the case where $R>>a$. This simplifies the algebra and moreover this is the case where (\ref{action}) reduces to (\ref{Sbeta}). This can then be compared with the result of the quadratic tachyon considered in section 5. The result is
\[
S=K{\p \over \p ln~a}[{{a\over 2R}({2R\over a})^{-k^2+2}\over (1-k^2)(2-k^2)}]
\]
\be
={({2R\over a})^{-k^2+1}\over {2-k^2}} \approx {R\over a}(1+(\hf -ln~{2R\over a})k^2+...)
\ee

We see that the relative coefficient of the $1$ and $k^2$ terms is critically dependent on $2R\over a$. We can choose it to match the derivative expansion obtained in section 5 but since this would not satisfy $R>>a$, we would have to go back to the general expression (\ref{S_B}). The conclusion is that the coefficient of $k^2$ term is not universal and depends on the regularization scheme. Thus the tachyon fields that correspond to different regularization are expected to be related by field redefinitions, but these could depend on $k$ :
\[
\phi (k) = \phi '(k) + A \al' k^2 \phi '(k) + ...~+O(\phi '(k)^2)+...
\]

{\bf \boldmath Normalization constant $K$ and Determination of $R/a$:}

 The function $h$ that multiplies the {\em on shell} kinetic term is universal in the sense that it multiplies the full action for {\em on-shell} tachyons, including the cubic term that is calculated in sec 4, and also the quartic and higher order terms, as argued in sec 4.2. Normally the S-matrix is supposed to capture the physics of a theory and to that extant we have a well defined physical theory. 

However the height of the tachyon potential is also in some sense "universal"  since it encodes information about the D-brane tension \cite{AS3}. This is related to the coefficient of the quadratic term of the {\em zero momentum} tachyon. This is quite a different function of $R/a$. Thus we can use this constraint in principle to determine the ration $R/a$ {\em within a regularization scheme}.
Also $K$ can be fixed once the string coupling is specified \cite{GS}. However if there are other such universal quantities, that would be a problem because  there are no further free parameters to fix. In such a situation more minute details of the regularization would have to be specified. This seems to be an open question at this stage.

\subsection{Renormalized Fields: Intermediate Case}

Working with bare fields we see that the continuum limit cannot be taken. Although there is no problem with this, it is interesting to consider other possibilities. For instance it is possible to absorb the $a$ dependence by renormalizing the tachyon field. In that case we can take
the limit $a \rightarrow 0$ without too much difficulty. In fact we can do it in two stages. We can define an intermediate field $\phi _I$ that absorbs the $a$ dependence that arises from normal ordering the vertex operators. The calculation is then very similar to the S-matrix calculation. We can also define a fully renormalized field $\phi _R$ in terms of which we can write the action without any $a$ dependence. In place of $a$ we have
a renormalization length scale $L$ entering the action. And the off-shell vertices will be functions of $L\over 2R$.

{\bf \boldmath $\phi _I$ (Intermediate Field):}

Define $\phi _I$ by:
\[
\int {dx\over a} ~\phi _B (k) \e = \int {dx \over a}~ \phi _B (k) ({a\over 2R})^{k^2\over 2}:\e : = 
\int {dx\over L} ~Z_1 \phi _I (k) \e 
\]
\be 
=\int {dx \over L}~ \phi _I (k) Z_1({a\over 2R})^{k^2\over 2}:\e :
\ee
Choose $Z_1=  ({L\over a})^{k^2\over 2}$.  Thus
\be  \label{phI}
\phi _B (k) = \phi _I (k) ({L\over a})^{{k^2\over 2}-1}
\ee
and
\be
\int {dx\over a} ~\phi _B (k) \e=\int {dx\over L}~\phi _I (k) ({L\over 2R})^{k^2\over 2} :\e :
\ee
Off-shell, when $a\rightarrow 0$ $\phi _I$ is a good variable to work with because it is finite. In fact if this is the entire $a$ dependence in the theory (i.e. if we ignore the nonlinearities) we get using (\ref{phI}),
\be
{d\over d~ln~a} \phi _I =0 = {d\phi _B \over d~ln~a} ({L\over a})^{1-{k^2\over 2}} + ({k^2\over 2}-1) \phi _B ({L\over a})^{1-{k^2\over 2}}
\ee
\be  \label{beta0}
\beta _{\phi _B} = ({k^2\over 2}-1) \phi _B
\ee

Let us use $\phi _I$ to calculate the action. The quadratic term is the same as (\ref{3.4}) with the replacements: $\phi _B \rightarrow \phi _I$ and $a \rightarrow L$.
\be
\int dk \int _{-R+a}^{R} {dz \over L} ~\int   _{-R} ^{z-a}{du \over L}
({z-u\over 2R})^{-k^2}({L\over 2R}) ^{k^2} \phi _I (k)  \phi _I (-k)
\ee
\be
=\int dk~{({L\over a})^{k^2-2}\over k^2-1} [{2R\over a}-1 - [ {({2R\over a})^{-k^2+2}-1\over -k^2+2}]]\phi _I(k)\phi _I(-k)
\ee

For near on shell fields with $k^2\rightarrow 2$, this becomes:
\be
{1\over k^2-1}[{2R\over a}-1 -ln~{2R\over a} -\hf (2-k^2) ln^2~{2R\over a}][1+(k^2-2) ln~{L\over a}]
\ee
For $k^2=2$ this becomes as before ${2R\over a} -1 -ln~{2R\over a}$. 

Then the actions becomes
\be
S= K{\p \over \p ln~a}\int dk~[1+ (k^2-2) ln~{L\over a} -\hf {(2-k^2) ln^2~{2R\over a} \over {2R\over a}-1 -ln~{2R\over a}}]
\phi _I(k) \phi _I (-k)
\ee

The limit $a \rightarrow 0$ is well defined and gives
\be
S= K\int dk~(2-k^2)\phi _I(k)\phi _I(-k)
\ee
For general $k^2$ we have 

\be   \label{S_I}
S=K \int dk~{\p\over \p~ln~a} {{({L\over a})^{k^2-2}\over k^2-1} [{2R\over a}-1 - [ {({2R\over a})^{-k^2+2}-1\over -k^2+2}]]\over {2R\over a}-1 -ln~{2R\over a}}\phi _I(k)\phi _I(-k)
\ee
In this expression, of course $a$ has to be kept finite, because we have not done a complete renormalization.

The connection between the two expressions can be seen as follows:
Let $Z(a, \phi _B)$ be the partition function in terms of bare fields. When we rewrite it in terms of
$\phi _I$ we get a function $Z_1$ that has a different {\em explicit} '$a$' dependence.
\be 
Z(a, \phi _B) = Z_1 (a, \phi _I (\phi _B,a))
\ee
Then
\be
{\p Z\over \p~ln~a} = {\p Z_1\over \p~ln~a} + {\p Z_1 \over \p \phi _I} {\p \phi _I \over \p ln~a}
\ee
Using ${d\phi _I \over d~ln ~ a}=0={\p \phi _I \over \p \phi _B}\beta _{\phi _B} + {\p \phi _I \over \p ln~a}$
we get
\be
{\p Z\over \p~ln~a}={\p Z_1\over \p~ln~a} -{\p Z_1 \over \p \phi _I}{\p \phi _I \over \p \phi _B}\beta _{\phi _B}
\ee

We see that on-shell, when $\beta_{\phi_B}=0$, the two actions are the same.
Using (\ref{phI}) and (\ref{beta0}) we get
\be   \label{329}
{\p Z\over \p~ln~a}={\p Z_1\over \p~ln~a} -{\p Z_1 \over \p \phi _I}({k^2-2\over 2})\phi_I
\ee

Comparing (\ref{S_B}) and (\ref{S_I}), we can check that they are related by (\ref{329}).

\subsection{Renormalized Fields: Full Renormalization}

{\bf \boldmath $\phi _R$ Renormalized Field:}

Let us now turn to the fully renormalized calculation:
\be
\int {dx\over a} ~\phi _B (k) \e=\int {dx\over L}~\phi _I (k) ({L\over 2R})^{k^2\over 2} :\e :
\ee
We write $\phi _I (k) = \phi _R(k) + \delta \phi _R (k)$. Thus
\be
\int {dx\over a} ~\phi _B (k) \e=\int {dx\over L}~\phi _R (k) ({L\over 2R})^{k^2\over 2} :\e :+\int {dx\over L}~\delta \phi _R (k) ({L\over 2R})^{k^2\over 2} :\e :
\ee

\subsubsection{Renormalization :}

$\delta \phi _R$ is a counterterm that will be chosen order by order in such a way that correlation functions are finite in the limit $a\rightarrow 0$. Of course this can only be achieved in some ranges of values of $k$ because in general there will be divergences corresponding to higher dimensional vertex operators. For general $k$ one will need all the massive mode operators as well which will be tantamount to an exact RG treatment of the problem of open strings.

The normal ordering divergences have been taken care of. So the first new divergence occurs in second order correction. Thus in any correlation function one can insert two powers of $\int {dx\over L}~\phi _R (k) ({L\over 2R})^{k^2\over 2}:\e :$:
\[
\langle O_1 O_2...\{\int dk_1~\int dk_2~\hf \int {dx_1\over L}~\phi _R (k_1) ({L\over 2R})^{k_1^2\over 2}:e^{ik_1X(x_1)} :\int {dx_2\over L}~\phi _R (k_2) ({L\over 2R})^{k_2^2\over 2}:e^{ik_2X(x_2)} :+ 
\]
\be
\int dk~\int {dx\over L}~\delta \phi _R (k) :e^{ikX(x)}:({L\over 2R})^{k^2\over 2} \}
.....O_N \rangle
\ee
$\delta \phi _R$ has to be chosen to make this finite. To quadratic order in $\phi _R$ and to zeroeth order in all other fields we can do this by removing the divergence associated with the contraction between $X(x_1)$ and $X(x_2)$ and subsequent integration over the positions. This can be summarized by using the OPE:
\[
\langle O_1 O_2...\{\int dk_1~\int dk_2~\hf \int {dx_1\over L}~ \int {dx_2\over L}~\phi _R (k_1)\phi _R (k_2) ({L\over 2R})^{k_1^2 +k_2^2\over 2}\Big\|{x_1-x_2\over 2R}\Big\|^{k_1.k_2}:e^{i(k_1X(x_1)+k_2X(x_2)} :-
\]
\be    \label{ope}
\int dk~\int {dx\over L}~\delta \phi _R (k) :e^{ikX(x)}:({L\over 2R})^{k^2\over 2} \}
.....O_N \rangle
\ee

To proceed further we need to specify some range of momenta. We consider two cases:

\subsubsection{\boldmath$ k_1 +k_2 = 0; k_1^2=k_2^2=2$} 

This is actually the calculation done earlier - when
$k_1+k_2$ is zero this is a contribution to the partition function. (i.e. without any operators $O_{1,2,3}$ in 
(\ref{ope}). However we will be able to take $a\rightarrow 0$  because the counterterms
remove all divergences - in this momentum range. This simplifies the results somewhat. 
Thus \footnote{Note that we are not renormalizing the one-particle irreducible effective action. We are interested in the full partition function, which is to be made finite. Therefore we do not subtract from the correlator the term $1+ln~({x_1-x_2\over 2R})$ as was done for instance in the original papers on the beta function method, for e.g. \cite{DS}.}
\be
-\int dk_1~\hf \int _{-\infty}^\infty {dx_1\over L}~ \int _{-\infty}^\infty {dx_2\over L}~\phi _R (k_1)\phi _R (-k_1) ({L\over 2R})^{k_1^2}\big\|{x_1-x_2\over 2R}\big\|^{-k_1^2}
\ee
\be
=-\int dk_1~\int {dx_1\over L}~ ~\phi _R (k_1)\phi _R (-k_1) (L)^{k_1^2-1}{(x_1-x_2)^{-k_1^2+1}
\big\|_{-R}^{x_1-a}\over -k_1^2+1}
\ee
\be
=-\int dk_1~\int _{-R+a}^{+R}{dx_1\over L}~ ~\phi _R (k_1)\phi _R (-k_1) (L)^{k_1^2-1}
{[a^{-k_1^2+1}- (x_1+R)^{-k_1^2+1}]\over -k_1^2+1}
\ee
\be
= -\int dk_1~{1\over -k_1^2+1}[({2R-a\over L})({a\over L})^{-k_1^2+1} - 
[{(2R)^{-k_1^2 +2}-(a)^{-k_1^2+2}\over -k_1^2+2}]L^{k_1^2-2}]\phi _R (k_1)\phi _R (-k_1)
\ee
\be
= -\int dk_1~{1\over -k_1^2+1}[({2R-a\over L})({a\over L})^{-k_1^2+1} - 
[{({2R\over L})^{-k_1^2 +2}-({a\over L})^{-k_1^2+2}\over -k_1^2+2}]]\phi _R (k_1)\phi _R (-k_1)
\ee

We add a counterterm $\int _{-R}^R {dx\over L} \delta \phi _R(0)= {2R\over L}\delta \phi _R(0)$, whose contribution to $Z$ is $\langle -{2R\over L} \delta \phi _R(0)\rangle$.
Choose
\[
\delta \phi_R(0) = -\int dk_1~{1\over -k_1^2+1}[(1-{a\over 2R})({a\over L})^{-k_1^2+1} + 
[{{L\over 2R}[ ({2R\over L})^{-k^2+2}-({a\over L})^{-k_1^2+2}]\over -k_1^2+2}]] \phi _R (k_1)\phi _R (-k_1) +
\]
\be
g \phi _R (k_1)\phi _R (-k_1)
\ee
We choose $g$ such that $\delta \phi _R(0) =0$ when $L=a$. This fixes 
\be    \label{g}
g=\{{1- {L\over 2R}\over -k_1^2+1} - ({L\over 2R})[{[({2R\over L})^{-k^2+2}-1]{1\over -k_1^2+2}\over -k_1^2+1}]\} 
\ee

So that the contribution to the partition function at this order is:
\be
-\int dk_1~ {2R\over L} g(k_1, {L\over 2R})\phi _R (k_1)\phi _R (-k_1)
\ee

Including the higher orders of the two point (i.e. two vertex operator)  functions has the effect of exponentiation: Thus $\phi _R(0) \rightarrow \phi _R(0) + g\phi _R (k_1)\phi _R (-k_1)$. The cubic and higher order correlations will introduce further modification. Another way to see this is that we can define an effective Lagrangian with $\phi _{eff}=\phi _R(0) +g\int dk_1~\phi _R (k_1)\phi _R (-k_1)$ as the coefficient of $1$. Thus 
\be   \label{Z}
Z = e^{-{2R\over L} \phi _R(0) - \int dk_1~{2R\over L} g(k_1, {L\over 2R})\phi _R (k_1)\phi _R (-k_1) }
\ee

{\bf Beta function (\boldmath $\beta$):}

Let us calculate the $\beta $ function to this order:
\[
{\phi _B (0)\over a} = {\phi _R (0)\over L} -{1\over L}  \{\int dk_1~{1\over -k_1^2+1}[(1-{a\over 2R})({a\over L})^{-k_1^2+1} +
\]
\be
[{{L\over 2R}[({L\over 2R})^{-k^2+2}-  ({a\over L})^{-k_1^2+2}]\over -k_1^2+2}]]\phi _R (k_1)\phi _R (-k_1)\}
 +{g\over L} \phi_R(k)\phi _R(-k)
\ee
We define $\beta _{\phi _R} = - {d\over dln~L} \phi _R$.  Note that the total power of $L$ in each part of the second order contribution is $k^2-2$. Thus operating with $\p\over \p ln~L$ just gives an overall multiplicative  factor of $k^2-2$. When we act with $d\over dln~L$ on $\phi _R(k)\phi _R (-k)$ we pick up a multiplicative factor
of $2-k^2$. This is because the beta function to lowest order for $\phi _R (k)$ is (as we will see below) 
is $\beta _{\phi _R(k)} = (k^2-2)\phi _R (k)$. These two contributions thus exactly cancel and we are left with the contribution from the term $g\phi _R(k) \phi _R(-k)$. 
\be
0=
{-\beta _{\phi _R(0)} -\phi _R(0)\over L} +  \int dk_1~{d\over dln~L}[({g\over L}) \phi _R(k_1)\phi _R(-k_1)] 
\ee

Thus
\be  \label{3.53}
\beta _{\phi _R (0)} = -\phi _R (0) + \int dk~[L{d\over d ln ~L} ({g\over L}) + g (2-k^2)]\phi _R (k) \phi _R(-k)
\ee

Note that the $a$ dependent terms disappear from the $\beta$ function.

Note also that the finite part of the counterterm makes a non trivial contribution to the $\beta$ function - just as it did to the partition function (\ref{Z}). This is to be expected. The finite term modifies the renormalization scheme, which in turn corresponds to field/coupling constant redefinitions. This is reflected in some redefinitions of the tachyon field also.  

Finally if we set $g=\{{1- {L\over 2R}\over -k_1^2+1} - ({L\over 2R})[{[({2R\over L})^{-k^2+2}-1]{1\over -k_1^2+2}\over -k_1^2+1}]\} $ from (\ref{g})
we find that the beta function becomes
\be     \label{betaR}
\beta _{\phi _R(0)}=-\phi _R(0) + \int dk~ (1-{L\over 2R} )\phi _R (k) \phi _R(-k)
\ee
 \be
=\beta _{\phi _B (0)} ~~~~at~~L=a
\ee
and agrees with (\ref{betabare}) when $L=a$.
- the beta function calculated in terms of bare fields. This is because the counterterms vanish when $L=a$, i.e. the two definitions of $\phi$ agree when $L=a$. 

{\boldmath $k^2=2$:}

Again for $k^2=2$ we have to be more careful. The partition function to this order is 
\be
Z= -{2R\over L} \phi _R (0) + [{2R\over a} -1 - ln {2R\over a}] \phi _R (k) \phi _R(-k) -{2R\over L} \delta \phi _R (0)
\ee

We let
\be
\delta \phi _R (0) = ({L\over a} -{L\over 2R}- {L\over 2R} ln~{2R\over a})\phi _R (k) \phi _R (-k) + g \phi _R(k)\phi _R(-k)
\ee

Now choose $g({L\over 2R})$ such that $\delta \phi _R(0)=0$ when $L=a$. Thus

\be
g({a\over 2R})= -(1- {a\over 2R}-{a\over 2R} ln~{a\over 2R}) 
\ee
Thus 
\be
g({L\over 2R}) = -(1- {L\over 2R}-{L\over 2R} ln~{L\over 2R}) 
\ee

This gives for the partition function:
\be
Z=1 -{2R\over L} \phi _R (0) - \int dk~{2R\over L} g\phi _R (k) \phi _R (-k)
\ee
or if we exponentiate due to the higher orders we get:
\be
Z=e^{-{2R\over L} \phi _R (0) - \int dk~{2R\over L} g\phi _R (k) \phi _R (-k)}
\ee

We can now calculate the beta function.
At $k^2=2$ since the leading contribution $({k^2\over 2}-1)\phi_R(k)$ vanishes, we get
\be
{d ({\phi _B (0)\over a})\over dln~L}=0=-{\beta _{\phi _R (0)}\over L} -{\phi _R(0)\over L} + {\p \over \p ln~L}({g\over L} )\phi _R(k)\phi _R(-k)
\ee
 which gives the same expression (\ref{betaR}) for the beta function and also agrees with the bare beta function.

We now turn to the second case:

\subsubsection{\boldmath$ (k_1 +k_2)^2 =  k_1^2=k_2^2=2$} 

In this case the operators are marginal and there are log divergences. The limits of integration,$ \pm R$ can be taken to be arbitrarily large and we can use translational invariance.
In this case $k_1.k_2 = -1$.  We go back to (\ref{ope})
\[
\langle O_1 O_2...\{\int dk_1~\int dk_2~\hf \int {dx_1\over L}~ \int {dx_2\over L}~\phi _R (k_1)\phi _R (k_2)
\]
\[ ({L\over 2R})^{k_1^2 +k_2^2\over 2}\Big\|{x_1-x_2\over 2R}\Big\|^{k_1.k_2}:e^{i(k_1X(x_1)+k_2X(x_2)} :- 
\]
\be    
\int dk~\int {dx\over L}~\delta \phi _R (k) :e^{ikX(x)}:({L\over 2R})^{k^2\over 2} \}
.....O_N \rangle
\ee
To leading order $e^{i(k_1X(x_1)+k_2X(x_2)}) \approx e^{i(k_1+k_2)X(x_1)}$. Using translational invariance and doing the integral over $x_2$ gives:
\be
\int dk_1~\int dk_2~ \int {dx_1\over L}~ ({L\over 2R})^{k_1^2+k_2^2\over 2} {1\over L} ({1\over 2R})^{k_1.k_2}{x^{k_1.k_2+1}\over k_1.k_2+1}\Big\| _a^R \phi _R (k_1)\phi _R (k_2)
\ee
\be
=\int dk_1~\int dk_2~ \int {dx_1\over L}~ ({L\over 2R})^{k_1^2+k_2^2\over 2} {1\over L} ({1\over 2R})^{k_1.k_2}R^{k_1.k_2+1}{[1-({a\over R})^{k_1.k_2 +1}]\over k_1.k_2+1}\Big\| _a^R \phi _R (k_1)\phi _R (k_2)
\ee
Since $k_1.k_2 +1\approx 0$ we can expand  in powers of $(k_1.k_2+1)ln~{a\over R}$. This gives (we set $k_1+k_2=k$)\footnote{ In all expressions replace $k_2$ by $k-k_1$.}
\be
-\int {dx_1\over L}\int dk~ \int dk_1~ ({L\over 2R})^{k^2\over 2} ({L\over R})^{-k_1.k_2 -1} ln~{a\over R} :e^{ikX}: 
\phi _R (k_1) \phi _R (k -k_1)
\ee

To cancel the divergence we choose a counterterm:
\be   \label{CT}
\delta \phi _R (k) = -\int dk_1~ [({L\over R})^{-k_1.k_2-1}ln~{a\over L} + f(k,k_1,{L\over R})] \phi _R (k_1)\phi _R (k-k_1)
\ee
We have added a finite term proportional to a function $f$. Thus the coefficient of the operator $e^{ikX}$ becomes modified by the addition of:
\be
-\int {dx_1\over L}\int dk~ \int dk_1~ ({L\over 2R})^{k^2\over 2}[ ({L\over R})^{-k_1.k_2 -1} ln~{L\over R} +f(k,k_1,{L\over R})] :e^{ikX}: 
\phi _R (k_1) \phi _R (k -k_1)
\ee

Note that as before if we want $\delta \phi _R(k)=0$ when $L=a$, $f$ is fixed to be zero.

{\bf Beta function\boldmath ($\beta$):}

We proceed now to calculate the $\beta$ function. Start with:

\be
{\phi _B (k) \over a} ({a\over 2R})^{k^2\over 2}= {\phi _R \over L} ({L\over 2R})^{k^2\over 2} + {\delta \phi _R \over L}({L\over 2R})^{k^2\over 2}
\ee

Apply ${d\over dln~a}$ to both sides to get:
\be   \label{352}
[-{\beta _{\phi _R(k)}\over L} + ({k^2\over 2}-1) {\phi _R(k)\over L}]({L\over 2R})^{k^2\over 2} +[({k^2\over 2}-1){\delta \phi_R(k)\over L} + {d\delta \phi_R(k)\over dln~L}{1\over L}]({L\over 2R})^{k^2\over 2}
\ee
Since $\delta \phi _R(k)$ is $O(\phi _R ^2)$, we have the lowest order result (which was used in (\ref{3.53}):
\be
\beta _{\phi _R(k)} = ({k^2\over 2}-1) \phi _R(k)
\ee
To get the next order contribution we need to use (\ref{CT}):
\[
{d\delta \phi _R (k)\over dln~L}=(1-{k^2\over 2})\delta \phi_R(k) +\int dk_1~({L\over R})^{-k_1.k_2-1}\phi _R (k_1)\phi _R(k_2) +
\]
\be   \label{354}
-\int dk_1 \{{df\over dln~L}  + f[(1-{k_1^2\over2}) + (1-{k_2^2\over2})]\}\phi _R(k_1)\phi _R(k_2)
\ee

Substituting (\ref{354}) in (\ref{352}), we see that the divergent terms proportional to $\delta \phi _R(k)$
cancel leaving us with:
\[
[-{\beta _{\phi _R(k)}\over L} + ({k^2\over 2}-1) {\phi _R(k)\over L}]({L\over 2R})^{k^2\over 2}  + \int dk_1~({L\over R})^{-k_1.k_2-1}({L\over 2R})^{k^2\over 2}\phi _R (k_1)\phi _R(k_2) +
\]
\be   \label{355}
-\int dk_1 \{{df\over dln~L}  + f[(1-{k_1^2\over2}) + (1-{k_2^2\over2})]\}\phi _R(k_1)\phi _R(k_2)
{1\over L}]({L\over 2R})^{k^2\over 2}
\ee
Thus:
\[
\beta _{\phi _R(k)}=({k^2\over 2}-1) \phi _R(k) +\int dk_1~({L\over R})^{-k_1.k_2-1}\phi _R (k_1)\phi _R(k_2) +
\]
\be   \label{beta}
-\int dk_1 \{{df\over dln~L}  + f[(1-{k_1^2\over2}) + (1-{k_2^2\over2})]\}\phi _R(k_1)\phi _R(k_2)
\ee
The second line in (\ref{beta}) represents the effect of the finite parts of the counterterm on the $\beta$ function.

\subsubsection{\bf Action}

Once we have $Z$ to second order, and also $\beta$, we can use (\ref{Sbeta}) to calculate the action.

\be   
S = K {\p \over \p ln~a}[{a\over 2r} Z] = K {a\over 2r}(1+\int dk~\beta _{\phi_B (k)}{\p \over \p \phi _B (k)})Z
\ee  
with
\be   
Z = e^{-{2R\over L} \phi _R(0) - \int dk_1~{2R\over L} g(k_1, {L\over 2R})\phi _R (k_1)\phi _R (-k_1) }
\ee
and
\[
\beta _{\phi _R(k)}=({k^2\over 2}-1) \phi _R(k) +\int dk_1~({L\over R})^{-k_1.k_2-1}\phi _R (k_1)\phi _R(k_2) 
\]
For concreteness  we take $g=-1$ (for on-shell fields, with $k^2=2$), and $f=0$. 
\be
S = K (1+ \int dk~ \beta _{\phi _R(k)}{\p \over \p \phi_R(k)})Z
\ee

Here, the symbol $\int dk$ stands for $\int _{-\infty}^{\infty} {d^{26}k\over (2\pi )^{26}}$. 

Using ${\p Z\over \p \phi _R(k)}={4R\over L} \phi_R(-k)$  we get:
\be
S= [1~+~{4R\over L} \phi_R(-k)({k^2\over 2}-1) \phi _R(k) + O(\phi ^3)]Z
\ee

This gives the kinetic term as expected. There is also a contribution of $O(\phi _R ^3)$ but since we need $Z$ to cubic order to get the full contribution we postpone this to the next section. The $\phi _R(0)$ contribution will also be treated separately. 

\section{Action for Tachyons with Generic Momenta:Cubic  Terms}

We now turn to a calculation of the cubic terms in the tachyon action. As in the last section we calculate it first using the bare Lagrangian. This can be done using the prescription (\ref{action}). 
Having calculated the action we can check whether the equation of motion is equal to the beta function to this (quadratic) order. Unfortunately the beta function has not been calculated for the bare theory to this order, only for the renormalized theory. However we have seen that for a 
particular choice of the counterterm we expect these to be equal at $L=a$. For on shell fields, which is what we treat in this section, this choice is $f=0$. Thus for this choice we know the beta function equation. We will see that it does coincide exactly with the equation of motion obtained from the tachyon action.
 
In principle one can work all this out in the renormalized theory. We will do the following. We show that for a particular choice of the cubic term in $Z$ (this uses the freedom of counterterms) the tachyon action gives the same equation as before (but in terms of renormalized fields) and hence the equation of motion agrees with the beta function - with $f=0$. What could be done further is to show by actual renormalization of the cubic term that this choice for $f$ is consistent with the choice of the  cubic term in $Z$. This is not  done in this paper.

\subsection{Bare Fields}

In order to keep track of the combinatorics in a transparent way we specialize the on-shell tachyon field to the following form
\[
\phi _B(X) = \phi_0(p) e^{ipX} +\phi_0(-p) e^{-ipX} +\phi_0(q) e^{iqX} +
\]
\be
\phi_0(-q) e^{-iqX} +
\phi_0(r) e^{irX} +\phi_0(-r) e^{-irX}
\ee
with some specific $p,q,r$ satisfying $ p=q+r$ and $p^2\approx q^2\approx r^2\approx 2$. Momentum conservation ensures that 
$\langle e^{ipX(z_1)} e^{-iqX(z_2)} e^{-irX(z_3)}\rangle$ and its complex conjugate expression are the only non zero three point correlators.
\subsubsection{Quadratic Term}

This has been calculated earlier. We are reproducing it here for the particular choice of the tachyon field.
The quadratic term in the partition function, $Z_2$, is 
\[
{1\over 2!}\int dz_1~\int dz_2~ [2\phi _0(p)\phi _0 (-p)\langle e^{ipX(z_1)}e^{-ipX(z_2)}\rangle +
\]
\be
2\phi _0 (q)\phi _0 (-q)\langle e^{iqX(z_1)}e^{-iqX(z_2)}\rangle+2\phi _0 (r)\phi _0 (-r)\langle e^{irX(z_1)}e^{-irX(z_2)}\rangle]
\ee

We use
\[
\int dz_1~\int dz_2~ \langle e^{ipX(z_1)}e^{-ipX(z_2)}\rangle = 2 \int _{-R+a}^{+R} dz_1~\int _{-R}^{z_2-a} dz_2~ \langle e^{ipX(z_1)}e^{-ipX(z_2)}\rangle
\]
\be
=2[{2R\over a} -1 -ln{2R\over a} +({p^2\over 2}-1) ln^2{2R\over a}]
\ee
Thus
\be
Z_2 = [\phi _0(p)\phi _0 (-p)2({2R\over a} -1 -ln{2R\over a} +({p^2\over 2}-1) ln^2{2R\over a})]+
[p\leftrightarrow q] + [p\leftrightarrow r]]
\ee
In terms of $\phi (k)$ this would be written as
\be
\hf \int dk~ \phi (k) \phi (-k)2[{2R\over a} -1 -ln{2R\over a} +({k^2\over 2}-1) ln^2{2R\over a}]
\ee

We are using a somewhat loose notation. Since we are considering a discrete set of momenta the $\int dk$ should actually be $\Sigma _{k=p,-p,q,-q,r,-r}$. The combinatoric factor $\hf$ outside should be noted.

\subsubsection{Cubic Term}

\be
{-1\over 3!}\int dz_1 ~\int dz_2~\int dz_3~ \Sigma _{k_1,k_2,k_3}\phi (k_1)\phi (k_2) \phi (k_3)
\langle e^{ik_1X(z_1)} e^{ik_2X(z_2)}e^{ik_3X(z_3)}\rangle
\ee
\[
=-\int dz_1 ~\int dz_2~\int dz_3~[\phi_0(p)\phi_0(-q)\phi _0(-r)\langle e^{ipX(z_1)}e^{-iqX(z_2)}e^{-irX(z_3)}\rangle +
\]
\be
\phi_0(-p)\phi_0(q)\phi _0(r)\langle e^{-ipX(z_1)}e^{iqX(z_2)}e^{irX(z_3)}\rangle]
\ee
We can restrict $z_3 <z_1$ without loss of generality. This gives a factor of two (actually it permutes the momenta)\footnote{The final answer for any term is independent of the momenta, so permuting momenta just gives a combinatoric factor} and then there are two inequivalent orderings: $z_1>z_2>z_3$ and $z_1>z_3>z_2$. Other orderings are related by Mobius transformations. So this gives another factor of two. 

The integral that needs to be done is of the form ($k_1=-p,k_2=q,k_3=r$):
\be
-({a\over 2R})^{k_1^2+k_2^2+k_3^2\over 2}\int _{-R+a}^{R} {dz_1\over a}~\int _{-R}^{z_1-a} {dz_3\over a}~|{z_3-z_1\over 2R}|^{k_1.k_3}\int _{z_3+a}^{z_1-a} {dz_2\over a}~|{z_2-z_3\over 2R}|^{k_2.k_3}|{z_1-z_2\over 2R}|^{k_1.k_2}
\ee
All the powers of $a,R$ cancel when the particles are on shell.
Make the change of variables: $z_2-z_3=y$ and the $y'={y\over z_1-z_3}$.
We get
\be
-\int _{-R+a}^{R} dz_1~\int _{-R}^{z_1-a} dz_3~|z_3-z_1|^{k_1.k_3+k_1.k_2+k_2.k_3+1}
\int _{a\over z_1-z_3}^{1-{a\over z_1-z_3}}dy'~|1-y'|^{k_1.k_2}y'^{k_2.k_3}
\ee
\[
=-\int _{-R+a}^{R} dz_1~\int _{-R}^{z_1-a} dz_3~|z_3-z_1|^{k_1.k_3+k_1.k_2+k_2.k_3+1}
\]
\be
[B(k_1.k_2+1,k_2.k_3+1)-B_{a\over z_1-z_3}(k_1.k_2+1,k_2.k_3+1)-B_{a\over z_1-z_3}(k_2.k_3+1,k_1k_2+1)]
\ee

Make the change of variables:$z_1-z_3=x$ and expand the incomplete Beta functions to get:
\[
=-\int _{-R+a}^{R} dz_1~\int _{2a}^{z_1+R} dx~|x|^{k_1.k_3+k_1.k_2+k_2.k_3+1}
\]
\be
[{1\over k_1.k_2+1}+ {1\over k_2.k_3+1}- {({a\over x})^{k_1.k_2+1}\over k_1.k_2+1}
-{({a\over x})^{k_2.k_3+1}\over k_2.k_3+1}]
\ee
Expanding in logarithms:
\[
=2\int _{-R+a}^{R} dz_1~\int _{a}^{z_1+R} dx~x^{-2}ln({x\over a})
\]
\be
=2[{2R\over a} -1 -ln{2R\over a} -\hf ln^2{2R\over a}]
\ee

Note that the precise choice of cutoff prescription reflected in the choice of the limits of integration affect the final result. This is as expected. This is equivalent to field redefinitions.

Thus we get
\be
4 \int _{-R}^{R} dz_1 \int _R^{z_1} dz_2~\int _{-R}^{z_2}dz_3~[\langle e^{-ipX(z_1)}e^{iqX(z_2)}e^{irX(z_3)}\rangle \phi_0(-p)\phi_0(q)\phi _0(r) +cc]
\ee

\be
=(4)2[{2R\over a} -1 -ln{2R\over a} -\hf ln^2{2R\over a}][\phi_0(-p)\phi_0(q)\phi _0(r)+cc]
\ee

Note that 
\[
\phi_0(-p)\phi_0(q)\phi _0(r)+cc={1\over 3!}\int dk_1~\int dk_2~ \int dk_3 ~\phi (k_1)\phi (k_2)\phi (k_3)\delta(k_1+k_2+k_3)
\]
Thus
\be
Z_3=8[{2R\over a} -1 -ln{2R\over a} -\hf ln^2{2R\over a}][\phi_0(-p)\phi_0(q)\phi _0(r)+cc]
\ee

Applying the prescription for the Action we get
\be
S=K{\p \over \p ln~a}[{Z_2+Z_3\over {2R\over a} -1 -ln{2R\over a}}]
\ee
We take the limit $R>>a$ to simplify things:
\[
=K\underbrace{{\p \over \p ln~a}  (2  {a\over 2R}ln^2{2R\over a})}_{2h({R\over a})} [\phi_0(p)({p^2\over 2}-1)\phi_0(-p) +p\leftrightarrow q + p\leftrightarrow r]+  
\]      
\be
  K \underbrace{{\p \over \p ln~a}(4{a\over 2R}ln^2{2R\over a})}_{4h({R\over a})}[\phi_0(p)\phi_0(-q)\phi_0(-r) +cc]
\ee
The function $h$ was introduced in Section 3.1.3.

The equation of motion obtained from $S$ is 
\be
{\p S \over \p \phi _0(-p)} =K2h[ ({p^2\over 2}-1)\phi_0(p)+2 \phi_0(q)\phi_0(r)]=0
\ee
This is to be compared with (\ref{beta}) that gives (for the choice $f=0$, which is when the bare and renormalized fields coincide at $L=a$, so we expect the beta functions to have the same form):
\[
\beta _{\phi_R(p)}=({p^2\over2}-1)\phi_0(p) +\int dk_1 \phi_R(k_1)\phi_R(p-k_1)
\]
\be
=({p^2\over 2}-1)\phi_0(p)+2 \phi_0(q)\phi_0(r)
\ee

\subsection{Quartic and Higher Order}

The calculation for higher order terms follows the same pattern. This has been known for a long time and has been used to obtain the tachyon equations to arbitrary order in perturbation theory for near on shell fields in \cite{BSPT}. It is known that when the particles are on shell then all N-point amplitude correlators have an SL(2,R) Mobius invariance. The result of integrating over $N-3$  positions has the form $1\over |x_1-x_2||x_2-x_3||x_3-x_1|$ $\times$ [effective action that generates S-matrix {\em minus} on shell poles].
The regularization of the integrals is what causes the subtraction of the divergent on-shell poles. Integrating over the remaining three is what gives the volume of the Mobius Group. (The cubic term we calculated above also has this form, when the particles are on shell.) Integrating over one of the three say $x_2$ gives the same integral as above:
\[
\int dx_1 ~\int dx_3~ {1\over |x_1-x_3|^2}ln {|x_1-x_3|\over a} 
\]
Thus the coefficient of the higher order terms in the action (for on shell fields) is the same as we found for the cubic term. Off shell it will be more of a more complicated form.

\subsection{Renormalized Fields}

The goal of this section is to outline how the calculation would be done in terms of renormalized fields. We have essentially all the ingredients necessary, except for the cubic term in the renormalized $Z$. For the purposes of this outline we will leave it as a free parameter. We show that it can be chosen by requirements of consistency. It needs to be checked by actual calculation that this term has this value.

The beta function has been calculated:
\be
\beta _{\phi _R (k)} = ({k^2\over 2}-1) \phi _R(k) + \int dk_1~ \phi _R(k_1) \phi _R(k-k_1)
\ee
\be
\Rightarrow \beta _{\phi_0(p)}=({p^2\over 2}-1)\phi_0(p) + 2 \phi _0(q) \phi _0(r)
\ee
\be
Z=[1+ 2(\phi _0(p)\phi_0(-p) +\phi _0(q)\phi_0(-q)+\phi _0(r)\phi_0(-r))+ b[\phi_0(p)\phi_0(-q)\phi_0(-r) +cc]]
\ee
The coefficient $b$ is the one that needs to be fixed by an actual calculation.

we calculate $Z+\Sigma _{k=p,q,r,-p,-q,-r} \beta_{\phi_0(k)}{\p Z\over \p \phi_0(k)}$.

We get
\[
S=K\{1+{8R\over L}[\phi_0(p)({p^2\over 2}-1)\phi_0(-p)+\phi_0(q)({q^2\over 2}-1)\phi_0(-q)+\phi_0(q)({q^2\over 2}-1)\phi_0(-q)]  +
\]
\be
 {24R\over L}[\phi_0(p)\phi_0(-q)\phi_0(-r) +cc] +b[\phi_0(p)\phi_0(-q)\phi_0(-r) +cc] + O(\phi_0^4)\}
\ee

The choice $b=-{8R\over L}$ yields the same action as we obtained earlier using the bare theory. As checked there the equations of motion agree with the beta function.

This concludes our discussion of the tachyon action with generic  momenta. We have seen that in the on-shell case it yields the expected space-time action. In the next section we turn to the (near) zero momentum case treated in \cite{WiI,WiII,LW,ShI,ShII,KMM}.

\section{Quadratic profile Tachyon and Tachyon Potential}

In this section we apply our prescription to the zero momentum tachyon, both in terms of bare fields as well as after renormalization. When $R>>a$ we have seen that this is the same as Witten's prescription so in this limit we recover the same results as \cite{WiII,ShI,KMM}. We will continue to work in the UHP as in the earlier sections, rather than on the unit disc. The leading terms (which is all we are interested in) are the same. We present the results here mainly for completeness and to emphasize that our prescription for the action agrees with earlier results when these are available.
\subsection{Bare Fields}

The starting point is the evaluation of the bare partition function:
\be
Z= \int {\cal D}X e^{-{1\over 2\alpha '}\int _\Gamma d^2x \p _\al X^i \p ^\al X^i -\int _{\p \Gamma} dx [{uX^2\over 2\al'}+\phi _0]  }
\ee
We assume $\Gamma$ is the Upper Half Plane (UHP) and $\p \Gamma$ is the boundary i.e. the Real axis. 
Then $\langle X^i (x)X^j(u)\rangle = -{\al'\over 2\pi}(ln |{x-u\over 2R}|- ln|{x-\bar u\over 2R}|)\delta ^{ij}$ is the Green function\footnote{The solution to $\p_\al \p^\al G(x,u)=\delta ^2(x-u),~~G(x,0,u,v)=0$ is ${1\over 2\pi}(ln |{x-u\over 2R}|- ln|{x-\bar u\over 2R}|)$.}
that satisfies Dirichlet boundary conditions $X(x,0)=0$.
The $\phi_0$ dependence of the partition function is trivially obtained: $\approx e^{-{2R\over a}}\phi_0$.
 In order to evaluate the $u$-dependence of the partition function we first evaluate the auxiliary quantity
\be
F[X_B(t)]=\int {\cal D}Xe^{-{1\over 2\alpha '}\int _\Gamma  d^2x \p _\al X^i \p ^\al X^i }\delta[X|_{\p \Gamma}-X_B(t)]
\ee
Let $X_c$ satisy  $\pp X_c=0$ and $X_c|_{\p\Gamma}=X_B(t)$. Then we write $X=X_c +x$ to get
\[ 
F[X_B(t)]=e^{-{1\over 2\alpha '}\int _\Gamma  d^2x \p _\al X_c^i \p ^\al X_c^i }\int {\cal D}xe^{-{1\over 2\alpha '}\int _\Gamma  d^2x \p _\al x^i \p ^\al x^i }\delta[x|_{\p \Gamma}]
\]
\[
=e^{-{1\over 2\alpha '}\int _\Gamma  d^2x \p _\al X_c^i \p ^\al X_c^i }
Det^{-D/2}[\pp ]
\]
\be
e^{-{1\over 2\alpha '}\int _{\p\Gamma}  dx  X_c(x,0)^i \p _y  X_c(x,0)^i }
Det^{-D/2}[\pp ]
\ee
For the problem at hand we can use the Green's function defined above to obtain
\be
X_c(x,y)=\int du \p_v G(x,y,u,v=0)X_B(u)
\ee
Thus
\be
F[X_B(t)]= e^{-{1\over 2\al '\pi} \int dx du {1\over (x-u)^2}X_B(x)X_B(u)} Det^{-D/2}[\pp ]
\ee
Introduce the Forier transform:
\[
 X_B(x) = 2R\int {dp\over 2\pi}~ e^{ipx}\tilde X_B(p)
\]
The factor of $2R$ has been added so that $X_B$ has the same scaling dimension (world sheet) as $X$.
\be
Z[u]=\int {\cal D}X_Be^{-{(2R)^2\over 2\al '} \int {dp\over 2\pi}(2|p|R+{2R\over a}u)\tilde X_B(pP\tilde X_B(-p)}
\ee
\be
=Det^{-D/2}[2R(|p|+{u\over a})] = e^{-{D\over 2}Tr~ln~[2R(|p|+{u\over a}]}
\ee
Now as long as $u\neq 0$ there is no infrared divergence. It is useful to separate out the zero mode.
Thus let us write
\[ 
\int _{-\infty} ^\infty {dp\over 2\pi} ln~ [2R(|p|+ {u\over a}]=ln {2Ru\over a} + 2\int _{2\pi \over 2R}^{2\pi\over a} ln~[2R(|p|+{u\over a}]
\]

The zero mode gives a prefactor $({a\over 2Ru})^{D/2}$. In the exponent we do the integral over $p$ and keep only terms to linear order in $u$. We take the ratio $Z[u]\over Z'[0]$, which gets rid of all the $u$-independent terms in Z[u]. The result is
\be
{Z[u,\phi _0]\over Z'[0]}=({a\over 2Ru})^{D/2} e^{-D{2R\over a}{u\over 2\pi} ln~{2R\over a}-{2R\over a}\phi_0}
\ee 
We can also trivially generalize to the case where there are $D$ u's i.e. $\Sigma _iu_i ({X^i})^2$ is the boundary Lagrangian density. We get:
\be
\bar Z \equiv {Z[u_i,\phi _0]\over Z'[0]}=[\Pi _{i=1}^D (\sqrt{a\over 2Ru_i})] e^{-\Sigma _{i=1}^D{2R\over a}{u_i\over 2\pi} ln~{2R\over a}-{2R\over a}\phi_0}
\ee 
Now we can apply our prescription for the action: 
\[
S =K {\p \over \p ln~a}\{ {a\over 2R} \bar Z \}
\]
\be
=K[1+\Sigma _i (\hf + {2R\over a}\phi_0 + {2R\over a} {u_i\over 2\pi} ln~{2R\over a} + {2R\over a} {u_i\over 2\pi}]\Pi _{i=1}^D (\sqrt{a\over 2Ru_i})] e^{-\Sigma _{i=1}^D{2R\over a}{u_i\over 2\pi} ln~{2R\over a}-{2R\over a}\phi_0}
\ee
Note that $D$ is 26 here (i.e. for the bosonic string). Also if we happen to set $2R=a$ the expressions simplify. While we do want to keep $a$ finite in general, for some purposes it is better to have the flexibility to tak $a\rightarrow 0$. In that case we need to renormalize the theory.

\subsection{Renormalized Fields}

We now renormalize the theory:
\[
S=\int dx~ [{\phi _{0B}\over a} + {u_B X^2\over 2\al' a}]
\]
\be
=\int dx~[{\phi _{0R}\over L} + {u_R X^2\over 2\al' L}]+[{\delta \phi _{0R}\over L} + {\delta u_R X^2\over 2\al' L}]
\ee

We evaluate the partition function perturbatively using the usual Green's function on the UHP
$\langle X(x)X(u)\rangle = -{\al'\over \pi}ln~{|x-u|\over 2R}$. 
The first term is ${u_R \over 2\pi L}ln~{a\over 2R}$. Choose $\delta \phi_{0R}={u_R \over 2\pi }ln~{a\over L}$ to make this finite:

\[\delta \phi_{0R} -{u_R \over 2\pi }ln~{a\over 2R}={u_R \over 2\pi }ln~{L\over 2R}
\]

At this order $\delta u_R =0$.

The relation between bare and renormalized parameters reads as:
\[
{\phi_{0B}\over a}= {\phi_{0R}\over L}+{\delta \phi_{0R}\over L}
\]
\[
\Rightarrow {\phi_{0B}\over a}= {\phi_{0R}\over L}-{u_R\over 2\pi L}ln~{L\over a}
\]
Also ${u_B\over a}={u_R\over L}$.
Using ${d\over d~ln~L}{u_B\over a} =0$ we get
\[
{-\beta _{u_R}\over L}-{u_R\over L} =0
\]
\[\Rightarrow \beta _{u_R} =- u_R\]
Also similarly
\[
{-\beta _{\phi_0}\over L}-{\phi_{0R}\over L} +{\beta _{U_R}\over 2\pi L}ln~{L\over a} + {u_R\over 2\pi L} ln~{L\over a} - {u_R\over 2\pi L} =0
\]
This gives:
\[
\beta _{\phi_{0R}}=-(\phi_{0R}+{u_R\over 2\pi})
\]
One can now write the partition function in terms of the renormalized quantities to obtain (for simplicity we are assuming $D=1$):
\[
Z=\sqrt{L\over u_R 2R} e^{-2R({\phi_{0R}\over L} + {u_R\over 2\pi L} ln~{2R\over L})}
\]
The formula for the action:
\[
S=K[1+\beta _{u_R}{\p\over \p u_R}+ \beta _{\phi_{0R}}{\p \over \p \phi_{0R}}]Z
\]
\[
=K[{3\over 2}+ {2R\over L} (\phi _{0R}+ {u_R\over 2\pi} + {u_R\over 2\pi} ln ~{2R\over L})]\sqrt{L\over u_R2R}e^{-{2R\over L} (\phi_{0R} + {u_R\over 2\pi} ln~{2R\over L})}
\]

Note that if $u=0$ to begin with, we trivially get for the tachyon potential (setting $L=2R$)
\[
V(\phi_0) = -K(1+\phi_0)e^{-\phi_0}
\]

The rest of the contribution can be understood as being due to the derivative terms in the action.
Thus we can match it with an effective action as was done in \cite{KMM}:
\[
S=K[e^{-\phi}[(1+\phi ) +{\al '\over 2\pi}\p_\mu \phi \p^\mu \phi]]
\] with $\phi = (\phi _0 + {uX^2\over 2\al '})$.

The potential term gives
\[
\int dX e^{-({uX^2\over 2\al '} + \phi _0)}(1+\phi _0 + {uX^2\over 2\al '})= e^{-\phi_0}({3\over 2}+\phi_0) \sqrt{2\pi\al '\over u})
\]
the kinetic term gives
\[
e^{-\phi_0}{u\over 2\pi}\sqrt{2\pi\al '\over u})
\]

The potential has a minimum at $\phi _0=\infty$ and the height of the potential is clearly equal
to the cosmological constant (or the brane tension). This is a partial verification of the Sen conjecture. Of course this does not fix the value of the brane tension in terms of the string coupling constant and the inverse string tension parameter $\al'$. As discussed in Sec 3.1.3, this is because the coefficient of the on shell cubic term is not fixed in terms of the cosmological constant  depending as it does on the details of the renormalization scheme. In particular there is a parameter $2R\over a$, even within this scheme.

\section{Summary and Conclusions}

The main aim of this paper is to work out the details of the prescription for a gauge invariant bosonic open string action that was presented in \cite{BSA}. An issue we discussed is the issue of renormalization. At least in certain ranges of momenta, where we can restrict our attention to just the tachyon sector, it is possible to renormalize the theory so that the cutoff parameter does not appear. This means that the continuum limit can be taken. There may be some conceptual attraction in doing this, but from a computational viewpoint nothing is gained. One conceptual advantage is that in the continuum limit one can make contact with another expression for the action derived in a completely different way in the  background independent formalism of Witten.

 The result is that both for on shell tachyons as well as for zero momentum tachyons the action can be constructed. The on shell action clearly is consistent with the S-matrix. We also showed that the equation of motion agrees with the $\beta$ function to this order.  For zero momentum tachyons the exact tree level potential can be calculated and agrees with earlier results using Witten's background independent formalism.

The normalization of the action depends on the renormalization scheme and introduces some arbitrariness. These have to be fixed by appealing to some universal quantities. One such is the 
S-matrix and another is the height of the tachyon potential. It is not clear whether there are  further constraints on the regularization scheme. This is an open question.

The issue of gauge invariance will arise when we deal with the vector and higher modes. The main obstacle in doing these calculations is the enormous amount of algebra. It would be useful if there are some clever field redefinitions that can be done to simplify things. 

The loop variable technique has been applied to closed strings at the level of equations of motion. If the open string action can similarly be generalized to closed strings one can tackle the question of closed string tachyon potential. Finally the connection with the exact RG needs to be understood better (both on the world sheet and in space-time). We hope to turn to these questions soon. 

{\bf Acknowledgements:} I thank Partha Mukhopadhyay for some useful discussions.

\end{document}